\documentclass[10pt]{iopart}
 \def\g{{\bf g}}
\def\k{{\bf k}}
\def\R{{\cal R}}
\def\qed{\vrule height8pt width6pt depth1pt}

\begin{document}
%{\hfill \tt DRAFT: \today}

\title[Cosmology, cohomology, and compactification]{Cosmology, cohomology, and compactification}

\author{C G Torre}

\address{Department of Physics, Utah State University, Logan, UT 84322-4415, USA}

\begin{abstract}

Ashtekar and Samuel have shown that Bianchi cosmological models with compact spatial sections must be of Bianchi class A.   Motivated by general results on the symmetry reduction of variational principles, we show how to extend  the  Ashtekar-Samuel results to the setting of weakly locally homogeneous spaces as defined, {\it e.g.}, by Singer and Thurston. In particular, it is shown that any $m$-dimensional homogeneous space $G/K$ admitting a $G$-invariant volume form will allow a compact discrete quotient only if the Lie algebra cohomology of $G$ relative to $K$ is non-vanishing at degree $m$.

\end{abstract}

Spatially homogeneous spacetimes have been studied extensively as both classical and quantum cosmological models (see, {\it e.g.,} \cite{MacCallum, Wiltshire}). In 3+1 dimensions all spatially homogeneous models but one (the Kantowski-Sachs model) admit a freely acting three-dimensional Lie group of isometries. These models are known as ``Bianchi models'' since they can be classified --- up to topology --- according to Bianchi's classification of three-dimensional Lie algebras. It was noted by Ashtekar and Samuel \cite{AS} that the orbit manifolds (the preferred spatial slices) for the Bianchi models can be compact only if the Lie algebra of the homogeneity group is Bianchi class A.  They also consider
a class of locally homogeneous geometries and show that the restriction to Bianchi class A Lie algebras is still necessary for compact spatial sections in this more general setting.  It is straightforward to generalize the Ashtekar-Samuel results to orbit manifolds of any dimension. The result is the same provided one generalizes ``Bianchi class A'' to ``unimodular''\footnote{A Lie group is said to be unimodular if it admits a bi-invariant measure. A Lie group is unimodular if and only if its adjoint representation on its Lie algebra is by linear transformations with determinant $\pm 1$. A Lie algebra is said to be unimodular if its adjoint representation is by trace-free linear transformations (in terms of structure constants, $C_{ab}^b=0$).  A connected Lie group is unimodular if and only if its Lie algebra is. For three-dimensional Lie algebras, unimodular is synonymous with Bianchi class A. }. However, to our knowledge it is unknown how to extend these results to the large class of ``weakly locally homogeneous spaces''\footnote{Unfortunately, the terminology here is not universal. Ashtekar and Samuel define ``globally homogeneous spaces''  the way  mathematicians normally define ``homogeneous spaces''. Further, Ashtekar and Samuel use ``homogeneous'' to denote a local form of homogeneity --- one that is inequivalent to what mathematicians normally mean by ``locally homogeneous'' and which Singer defines as ``weakly locally homogeneous''. We will use the latter term (in a no-doubt hopeless attempt) to minimize confusion.} of Singer, Thurston, {\it etc.} \cite{Singer, Thurston, Scott}. Such spaces need not satisfy the local homogeneity criteria of \cite{AS}.  For example, as pointed out in \cite{AS}, compact, weakly locally homogeneous spaces such as arising in the Kantowski-Sachs model, or, more generally,  such as  obtained by taking the product of a circle with a compact Riemann surface, are not covered by the results of \cite{AS}. The purpose of this note is to show how to extend the results of \cite{AS} to the weakly locally homogeneous setting.

The results of \cite{AS} on compactification of Bianchi models are closely related to the fact that the class A models obey the principle of symmetric criticality, that is, they always inherit variational principles by symmetry reduction \cite{Hawking, MT}. Indeed, the obstruction to the  spatially homogeneous symmetry reduction of, say, the Einstein-Hilbert action is a boundary term which trivially vanishes for compact spatial sections.  Recently, necessary and sufficient conditions for the validity of the principle of symmetric criticality were obtained for any isometry group action and for any metric theory of gravity \cite{FT}. These conditions include the unimodularity condition for Bianchi models, but in the general case they involve the relative cohomology of the symmetry Lie algebra. These results are easily extended to the (weakly) locally homogeneous setting.  This suggests that one can generalize the necessary condition of \cite{AS} for the existence of compact spatial slices to weakly locally homogeneous spaces in terms of relative Lie algebra cohomology.  In this note we show how to do this via an extension of the proof used in \cite{AS} for globally homogeneous spaces. In particular, we show that any $m$-dimensional homogeneous space $G/K$ (admitting a $G$-invariant volume form) allows a compact discrete quotient only if the Lie algebra cohomology of $G$ relative to $K$ does not vanish at degree $m$.  This result can be viewed as providing an extension of the results of \cite{AS} to the weakly locally homogeneous setting.

We begin by reviewing some definitions. A smooth manifold $M$ is a {\it homogeneous manifold} if it admits a transitive Lie group $G$ of diffeomorphisms.  The {\it isotropy group} $K\subset G$ of a point $p\in M$ is the subgroup which fixes $p$.  $M$ is diffeomorphic to the manifold of right cosets:  $M\approx G/K$.   With this identification, the transitive $G$ action on $M$ is the projection to $M$ of the left action of $G$ on itself.  If $M$ is equipped with a $G$-invariant Riemannian metric we say that $(M,g)$ is a {\it homogeneous geometry}. A generalization of a homogeneous manifold is a {\it Clifford-Klein manifold} $\bar M$, which is the quotient of $M$ by a discrete subgroup $\Gamma\subset G$  acting freely and properly discontinuously, $\bar M = M/\Gamma$.  The covering map from $M$ to $\bar M$ is denoted by $\bar\pi\colon M\to \bar M$. In general, $G$ does not drop to act on $\bar M$, although the Lie algebra of vector fields generating the action of $G$ on $M$ will drop to $\bar M$ locally.  A Riemannian metric $g$ on a manifold $N$ is called {\it weakly locally homogeneous} if given any two points $x,y\in N$ there exist isometric neighborhoods of $x$ and $y$.  We call such a pair $(g,N)$ a {\it weakly locally homogeneous geometry}. The manifold $N$ supporting a complete, weakly locally homogeneous metric can always be realized as a Clifford-Klein manifold $M/\Gamma$, with $(M,\bar\pi^*g)$ being a homogeneous geometry \cite{Scott, Kobayashi}. 

Given a Lie algebra $\g$ and a sub-algebra $\k$, {\it the Lie algebra cohomology of $\g$ relative to $\k$}, $H^*(\g,\k)$, is defined as the cohomology of forms on $\g$ invariant under the adjoint action of $\k$ on $\g$ \cite{Chev-Eil}. For our purposes, it is convenient to define this cohomology in terms of differential forms on a given Lie group $G$ with Lie algebra $\g$. Consider the vector space $\Omega^p(G)$ of left-invariant $p$-forms on $G$.  Let $K$ be a closed subgroup of $G$. Denote by $\Omega^p(G,K)\subset \Omega^p(G)$ the vector subspace of left $G$-invariant $p$-forms that are also right $K$-basic. A right $K$-basic form $\omega\in \Omega^p(G,K)$  is invariant under the right action of $K$ on $G$ and satisfies 
\begin{equation}
X\cdot \omega  = 0
\end{equation}
for all vector fields $X$ generating the right action of $K$ on $G$.  It is straightforward to verify that $d\colon \Omega^p(G,K)\to \Omega^{p+1}(G,K)$, whence one can define the relative Lie algebra cohomology $H^*(\g,K)$ as the cohomology in $\Omega^*(G,K)$. When $K$ is disconnected this is actually a slight generalization of the usual definition of relative Lie algebra cohomology $H^*(\g,\k)$.   If $K$ is connected then $H^*(\g,K)=H^*(\g,\k)$, that is, the relative cohomology  depends only on the Lie algebra data $\g$ and $\k$ and can be computed entirely in terms of the structure constants of $\g$.

Our main result is the following theorem which shows that the vanishing of a particular relative Lie algebra cohomology class is an obstruction to the existence of a compact Clifford-Klein manifold.

\bigskip
\noindent
{\bf Theorem.}
{\it Let $M\approx G/K$ be an $m$-dimensional homogeneous manifold admitting a $G$-invariant volume form. If $M$ admits a compact Clifford-Klein quotient $\bar M= M/\Gamma$  then $H^m(\g,K)\neq 0$.}

\bigskip\noindent
{\bf Proof:}

Let $\epsilon$ be a $G$-invariant volume form on $M$. Since $\Gamma\subset G$, there exists a volume form $\bar\epsilon$ on $\bar M$ defined by $\epsilon=\bar\pi^*\bar\epsilon$. If $\bar M$ is compact, then the integral of $\bar\epsilon$ over $\bar M$ is well-defined and 
\begin{equation}
\int_{\bar M} \bar\epsilon \neq 0.
\end{equation} 
Define $\nu\in \Omega^m(G,K)$ by $\nu = \pi^*\epsilon$, where $\pi$ is the surjection $\pi\colon G\to G/K$. Because $d\epsilon = 0$ we have that $d\nu =0$. Now suppose that $H^m(\g,K)=0$.  We then get $\nu = d\alpha$, where $\alpha\in \Omega^{m-1}(\g,K)$. This implies that $\pi^*(\epsilon - d\beta) = 0$, where $\beta$ is the $G$-invariant $(m-1)$-form defined by $\pi^*\beta = \alpha$.  Since $\beta$ is $G$-invariant and $\Gamma\subset G$, it follows that there exists a form $\bar\beta$ such that $\beta = \bar\pi^*\bar\beta$, whence $\bar\pi^*(\bar\epsilon - d\bar\beta)=0$. Thus $\bar\epsilon$ is exact, leading to
\begin{equation}
\int_{\bar M} \bar\epsilon = 0,
\end{equation}
which is a contradiction invalidating the assumption that $H^m(\g,K)=0$.  \qed
\bigskip

The requirement that $G/K$ admits a $G$-invariant volume form is equivalent to the requirement that the adjoint representation of $K$ on the vector space $\g/\k$ is by linear transformations with unit determinant.  In the case where $\bar M$ is orientable and supports a complete weakly locally homogeneous metric this requirement is automatically satisfied. In general, however, a locally homogeneous metric is not needed to establish this theorem ({\it c.f.} [AS]). In any case, if $K$ is connected this requirement is only on Lie algebraic data, {\it i.e.}, on structure constants.  Thus, provided $K$ is connected, the obstruction to compactification specified by the above theorem is completely determined by the Lie algebras $\g$ and $\k$. 

Let us consider some applications of this Theorem. 

\bigskip\noindent
{\sl Example 1: $\bar M = G$} 
\medskip

Suppose that $\bar M$ is in fact a connected Lie group $G$. In this case the Lie algebra cohomology condition  is equivalent to the  unimodularity of $G$ ---  in agreement with \cite{AS}  (see also \cite{Milnor}). To see this, one simply sets $\Gamma = identity$; the relative cohomology condition is then equivalent to the non-vanishing of the top-degree Lie algebra cohomology, which is equivalent to unimodularity (see, {\it e.g.} \cite{AF}). 

\bigskip
Next let us consider spatial manifolds of the form $\bar M=S^1\times {\R}_g$, where $\R_g$ is a compact, orientable Riemann surface of genus $g$.  Each of these manifolds can be viewed as a compact Clifford-Klein manifold. Moreover, in each case the underlying homogeneous space  has an Abelian (SO(2)) isotropy group so the volume form hypothesis is satisfied (which can also be seen from the fact that the manifolds in question admit a weakly locally homogeneous metric).  Therefore the obstruction displayed in the theorem above should be absent.

\bigskip
\noindent
{\sl Example 2: $\bar M= S^1\times S^2$}
\medskip

Begin with the homogeneous manifold $M=R^1\times S^2\approx G/K$ where $G= R^1 \times SO(3)$ and $K=SO(2)\subset SO(3)$. This is the group action featuring in the  Kantowski-Sachs model. Choosing $\Gamma = {\bf Z}\subset R^1$ we have $\bar M = (R^1\times SO(3))/({\bf Z}\times SO(2))$.\footnote{{\bf Z} is the additive group of integers. In this example $\bar M$ is  a homogeneous manifold.}  A basis of left-invariant 1-forms on $G$ is $(\omega^i, \omega^4)$, $i=1,2,3$, where 
\begin{equation}
d\omega^i= -{1\over 2} \epsilon^i{}_{jk} \omega^j\wedge\omega^k,\quad d\omega^4=0.
\end{equation}
The dual basis is $(e_i,e_4)$ and the Lie algebra of $K$ can be taken to be spanned by $e_3$. The vector spaces of left $G$-invariant, right $K$-basic 2-forms and 3-forms are spanned, respectively, by $\omega^1\wedge\omega^2$, and $\omega^1\wedge\omega^2\wedge\omega^4$, which satisfy
\begin{equation}
d(\omega^1\wedge\omega^2\wedge\omega^4) = 0,\quad d(\omega^1\wedge\omega^2)=0,
\end{equation}
so that $H^3(\g,K)=H^3(\g,\k)=R^1$.

\bigskip\noindent
{\sl Example 3: $\bar M = S^1\times T^2$}
\medskip

Set $G=R^1\times E(2)$, where $E(2)$ is the Euclidean group acting on the plane. $G$ acts transitively on $M=R^1\times R^2$ in the obvious way.  The isotropy group of the origin is $SO(2)\subset E(2)$ and we have $M = R^1\times E(2)/SO(2)$. We set $\Gamma = {\bf Z} \times {\bf L}$, where ${\bf Z}\subset R^1$ and ${\bf L}\subset E^2$ is a lattice in $R^2$. We can then realize $\bar M =S^1\times T^2= M/ \Gamma $. Note that, in general, $\bar M$ is neither a group manifold nor a homogeneous manifold.   A basis of left-invariant 1-forms on $G$ is given by $\omega^i$, $i=1,2,3,4$ where
\begin{equation}
d\omega^1 = -\omega^2\wedge\omega^3,\quad
d\omega^2 = \omega^1\wedge\omega^3,
\quad d\omega^3=0=d\omega^4.
\end{equation}
Denoting the dual basis by $e_i$, the isotropy sub-algebra is spanned by $e_3$. The vector space of left $G$-invariant, right $K$-basic 3-forms  is spanned by $ \omega^1\wedge\omega^2\wedge \omega^4$, while the space of left $G$-invariant, right $K$-basic 2-forms is spanned by $\omega^1\wedge\omega^2$. Both of these forms are closed, whence $H^3(\g,K)=H^3(\g,\k)=R^1$.

\bigskip\noindent
{\sl Example 4: $\bar M = S^1\times \R_g$, $g\geq 2$}
\medskip

Here $\R_g$ is a compact Riemann surface of genus $g\geq 2$,  {\it i.e.}, a $g$-handled torus.  We choose $G=R^1\times PSL(2,R)$, where $R^1$ is the additive group of reals,  and $PSL(2,R)=SL(2,R)/\{\pm Identity\}$. $G$ acts transitively on $R^1\times U^2$, where $U^2=\{(x,y)|y>0\}$ is the upper-half plane. In this group action $PSL(2,R)$ acts transitively on $U^2$ by linear fractional transformations.  The isotropy group of a point is easily seen to be an $SO(2)$ subgroup of $PSL(2,R)$, whence $M=R^1\times U^2\approx R^1\times PSL(2,R)/SO(2)$. For each choice of $\R_g$, there is a free, discrete subgroup $\gamma\subset PSL(2,R)$ such that $S^1\times \R_g \approx R^1\times U^2/({\bf Z}\times \gamma)$, {\it i.e.}, $\Gamma={\bf Z}\times \gamma$  (see, {\it e.g.} \cite{FK}).    The Lie algebra of $R^1\times PSL(2,R)$ can be defined by the basis $e_i$, $i=1,2,3,4$ satisfying
\begin{equation}
[e_1,e_2] = e_1,\quad [e_1,e_3] = -2 e_2,\quad [e_2,e_3] = e_3,\quad [e_i,e_4] = 0,
\end{equation}
with dual basis $\omega^i$ satisfying
\begin{equation}
\eqalign{
&d\omega^1 = -\omega^1\wedge\omega^2,\quad d\omega^2 = 2 \omega^1\wedge\omega^3,
\quad d\omega^3 = -\omega^2\wedge\omega^3,\cr 
&d\omega^4 = 0.}
\end{equation}
The isotropy subgroup $SO(2)$ of the point $(0,0,1)\in R^1\times U^2$ is generated by $e_1-e_3$. The set of $G$-invariant, $SO(2)$-basic 3-forms is spanned by the closed form $(\omega^1 + \omega^3)\wedge\omega^2\wedge\omega^4$. The set of $G$-invariant,  $SO(2)$-basic 2-forms is spanned by the closed form $(\omega^1 + \omega^3)\wedge\omega^2$, whence $H^3(\g,K)=H^3(\g,\k)=R^1$. 

\bigskip\noindent{\sl Example 5: A non-compactifiable LRS manifold}
\medskip

As our final example we exhibit a homogeneous manifold with local rotation symmetry that exhibits the Lie algebra cohomology obstruction to existence of a compact discrete quotient.  The transformation group $G$ is four-dimensional with coordinates $\lambda^\alpha$, $\alpha = 1,2,3,4$. In coordinates $x^i$, $i=1,2,3$, on $M=R^3$ we define the transitive $G$-action
\begin{eqnarray}
x^1\longrightarrow &x^1-\lambda^3,\\
x^2\longrightarrow &\lambda^1 + e^{\lambda^3}[x^2\cos(\lambda^4) - x^3\sin(\lambda^4)],\\
x^3 \longrightarrow    &\lambda^2 + e^{\lambda^3}[x^3\cos(\lambda^4) + x^2\sin(\lambda^4)].
\end{eqnarray}
The infinitesimal generators of this group action are
\begin{equation}
\eqalign{
&e_1 = \partial_2,\quad e_2=\partial_3,\quad e_3=-\partial_1 + x^2\partial_2 + x^3\partial_3,\cr &e_4=x^2\partial_3 - x^3\partial_2,}
\label{gen5}
\end{equation}
giving the Lie algebra
\begin{equation}
\eqalign{
&[e_1,e_2]=0,\quad[e_1,e_3] = e_1,\quad [e_2,e_3] = e_2,\cr
&[e_1,e_4]=e_2,\quad [e_2,e_4] = -e_1,\quad [e_3,e_4] = 0.}
\label{Lie5}
\end{equation}
The   isotropy subgroup of the origin is $K=SO(2)$ and is generated by $e_4$ (so that $M$ admits a $G$-invariant volume form as well as a $G$-invariant metric).    Denoting the dual basis for the Lie algebra as $\omega^\alpha$, $\alpha = 1,2,3,4$, the vector space of left $G$-invariant, right $SO(2)$ basic 3-forms on $G$ is spanned by $\omega^1\wedge\omega^2\wedge\omega^3$, which satisfies
\begin{equation}
\omega^1\wedge\omega^2\wedge\omega^3 = d(\omega^1\wedge\omega^2),
\end{equation}
where $\omega^1\wedge\omega^2$ is a left $G$-invariant, right $SO(2)$ basic 2-form. Thus $H^3(\g,\k)=0$ and $M$ does not admit a compact discrete quotient.  Every 
four-dimensional group action that (i) is transitive on a 3-manifold, (ii) admits a $G$-invariant Riemannian metric, and (iii) realizes the Lie algebra cohomology obstruction to a compact discrete quotient has the Lie algebra (\ref{Lie5}) and generators given (locally) in the form (\ref{gen5}).

We remark that in each of examples 4 and 5 the group action admits a transitive 3-dimensional subgroup with Lie algebra of Bianchi class B.  

\ack

This work was supported in part by National Science Foundation grant PHY-0244765 to Utah State University.

\end{document}